# Engineering the electronic and optical properties of 2D porphyrin-paddlewheel metal-organic frameworks

Victor Posligua,[1] Dimpy Pandya,[1] Alex Aziz,[2] Miguel Rivera,[2] Rachel Crespo-Otero,[2] Said Hamad,[3] Ricardo Grau-Crespo[1*]

[1]*Department of Chemistry, University of Reading, Whiteknights, Reading RG6 6AD, United Kingdom.*

*Email: r.grau-crespo@reading.ac.uk*

[2]*School of Biological and Chemical Sciences, Queen Mary University of London, Mile End Road, London E1 4NS, United Kingdom.*

[3]*Department of Physical, Chemical and Natural Systems, Universidad Pablo de Olavide, Ctra.de Utrera km.1, 41013 Seville, Spain.*

**Abstract**

Metal-organic frameworks (MOFs) are promising photocatalytic materials due to their high surface area and tuneability of their electronic structure. We discuss here how to engineer the band structures and optical properties of a family of two-dimensional (2D) porphyrin-based MOFs, consisting of *M*-tetrakis(4-carboxyphenyl)porphyrin structures (*M*-TCPP, where *M* = $Zn^{2+}$ or $Co^{2+}$) and metal ($Co^{2+}$, $Ni^{2+}$, $Cu^{2+}$ or $Zn^{2+}$) paddlewheel clusters, with the aim of optimising their photocatalytic behaviour in solar fuel synthesis reactions (water-splitting and/or $CO_2$ reduction). Based on density functional theory (DFT) and time-dependent DFT simulations with a hybrid functional, we studied three types of composition/structural modifications: a) varying the metal centre at the paddlewheel or at the porphyrin centre to modify the band alignment; b) partially reducing the porphyrin unit to chlorin, which leads to stronger absorption of visible light; and c) substituting the benzene bridging between the porphyrin and paddlewheel, by ethyne or butadiyne bridges, with the aim of modifying the linker to metal charge transfer behaviour. Our work offers new insights on how to improve the photocatalytic behaviour of porphyrin- and paddlewheel-based MOFs.

# 1. Introduction

Metal-organic frameworks (MOFs) are materials where metal atoms or clusters are connected via organic linkers to form rigid frameworks, often with a porous structure [1]. They have found applications in gas (*e.g.* $CO_2$, $H_2$) storage and separation [2-6]. More recently, MOFs have been investigated as potential photocatalysts due to their tuneable electronic structure and optical behaviour, as there is a huge compositional space of transition metal ions/clusters or multidentate organic linkers that can be incorporated into the framework [7-9]. Several strategies have been designed to modify MOFs for enhanced photocatalytic performance, including linker functionalization [10-12], mixing metals/linkers [13-16], and metal nanoparticle loading [17-20].

Porphyrin-like organic units are particularly attractive as components of MOF photocatalysts, due to their remarkable light absorbing properties, which is also the basis of natural photosynthetic systems [21-23]. Several porphyrin-based MOFs have been investigated for photocatalysis. For example, Fateeva *et al.* reported a water-stable porphyrin-based MOF with Al-carboxylate clusters as metal nodes, capable of performing photocatalytic production of hydrogen from water, in the presence of Pt nanoparticles [19]. A MOF consisting of Zn metalloporphyrins connected to $Zr_6O_8$ clusters through carboxylic groups, coupled with an organometallic [$Fe_2S_2$] complex, has also shown photocatalytic activity for hydrogen evolution [24]. Leng *et al.* [25] reported an indium-based porphyrinic MOF, USTC-8(In), where one-dimension In−oxo chains are connected by the porphyrin units, with excellent photocatalytic $H_2$ production under visible light. In this system, the out-of-plane $In^{3+}$ ions detach from the porphyrin ligands under excitation, avoiding the fast back electron transfer and thus electron-hole separation is improved. MOFs consisting of $Ru_2$ paddlewheel units and Zn-porphyrins were reported by Lan *et al.* [26] to exhibit visible-light photocatalytic activity for hydrogen evolution, without the presence of metal co-catalysts. In that case, the proximity of the Ru cluster to the porphyrin (~11 Å) was found to facilitate the electron transfer from the photoexcited porphyrins to the metal clusters.

There has been recent interest in creating two-dimensional (2D) photocatalytic MOFs, which could benefit from very accessible active sites and short paths for the photogenerated charge carriers to reach the solid-water interface. Wang *et al.* [20] have proposed ultrathin porphyrin-based MOFs consisting of $Ti_7O_6$ clusters and free-based porphyrins connected by $H_2$TCPP linkers, which exhibited excellent photocatalytic hydrogen evolution in the presence of Pt as co-catalyst. Porphyrin-based quasi-2D lanthanide MOFs with different thicknesses where synthesised by Jiang *et al.* [27], demonstrating that the thinner materials had higher Brunauer–Emmett–Teller (BET) surface area, light harvesting ability, carrier density, separation efficiency, and therefore better photocatalytic performance.

Despite the remarkable progress in recent years, porphyrin-based MOFs still need efficiency improvement in the light absorption and charge separation processes to become viable photocatalysts. The optical behaviour of porphyrin-based MOFs is still not well understood at a fundamental level, which hinders the optimisation process. Computer simulations based on density functional theory (DFT) can be very useful in rationalising the electronic and optical properties of MOFs [28]. Previous DFT simulation work from our group [29, 30] on 3D porphyrin-based MOFs similar to those synthesised by Fateeva *et al.* [19] has shown that the choice of metal at the porphyrin centre or at the metal clusters can be used to optimise the band alignment for the photocatalytic process. Also, we showed that when Al cations in the PMOFs are replaced by Fe cations, the position of the conduction band edge is lowered significantly, and that the Fe/Al composition in a mixed metal system can be used to tune the band edge positions.

In this work, we investigate a class of 2D porphyrin-based MOFs consisting of *M*-tetrakis(4-carboxyphenyl)porphyrin structures (*M*-TCPP, where *M* = $Zn^{2+}$ or $Co^{2+}$) and metal paddlewheel clusters ($M_2(COO^-)_4$). It is known that the metal paddlewheel cluster structure can accommodate different metals, *e.g.* $M = Co^{2+}$, $Ni^{2+}$, $Cu^{2+}$, $Zn^{2+}$, $Cd^{2+}$, $Mn^{2+}$ [31]. We consider here metal paddlewheels made of late 3*d* metals ($M = Co^{2+}$, $Ni^{2+}$, $Cu^{2+}$, or $Zn^{2+}$). The paddlewheel cluster comprises of two divalent metal ions bridged together by four carboxylate ligands. Each paddlewheel is then linked to four porphyrin units (**Figure 1**). A MOF with this layer structure was synthesised by Choi *et al.* [32] in bulk form, with layers exhibiting AB stacking. Zhao *et al.* [33, 34] showed how to grow this material anisotropically, with the help of surfactants, to create 2D nanosheets with only 8±3 layers. Spoerke and co-workers [35] also reported the synthesis and characterisation of 2D structures consisting of Zn-paddlewheels and Zn-porphyrins and showed that these structures can serve as active component in photovoltaic systems.

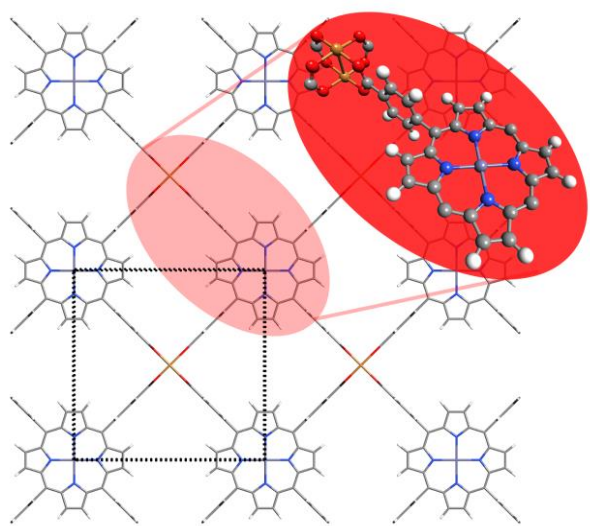

**Figure 1. Structure of *M*-Zn-TCPP nanosheet. Inside the red-shared circle, detailed structure of a single model of paddlewheel/linker/porphyrin conforming the MOF and dashed black lines show the unit cell. Colour code: grey = carbon, red = oxygen, white = hydrogen, blue = nitrogen, purple = zinc and orange = transition metal.**

The presence of porphyrin in a 2D pattern, and the possibility of tuning the electronic properties by changing the paddlewheel metal, make this family of materials potentially interesting for photocatalysis. One possible concern is the poor hydrolytic stability of some metal paddlewheel frameworks [36, 37]. At harsh environments (*i.e.* high loads of water), a hydrolysis reaction of water molecules with Cu-O-C group can induce the structural decomposition of Cu paddlewheel [36]. However, the stability in water depends on the nature of the metal in the paddlewheel and on the water loading [38, 39] which means that some of the 2D-MOFs investigated here could still be useful as photocatalysts. They are also promising in other applications (*e.g.* as photovoltaic materials), where the electronic properties are of interest.

Due to their simplicity, these 2D-MOFs provide useful model systems to investigate how to engineer the electronic and optical behaviour of porphyrin-based MOFs. The aim of this paper is to investigate how different modifications (change in metal centres, functionalization of the porphyrins, or changes in the organic bridge between the porphyrin and the paddlewheel) can tune the electronic and optical properties of 2D porphyrin-paddlewheel MOFs for photocatalytic and other applications.

## 2. Methodology

Calculations of the periodic models were performed using density functional theory (DFT) as implemented in the VASP program [40, 41]. The simulation cell consists of one $Zn^{2+}$ centred porphyrin linker and one metal paddlewheel node (**Figure 1**). Vacuum regions separating the layers from their periodic images have a width of ~20 Å.

The geometry of the atomic structure and lateral lattice constants were optimised using generalised gradient approximation (GGA) with Perdew-Burke-Ernzerhof (PBE) [42, 43] exchange-correlation functional. Hubbard corrections were applied for $Cu^{2+}$ and $Ni^{2+}$, using $U_{eff}$ values of 4.0 and 6.4 eV, respectively [44], and Grimme's empirical corrections were used to properly account for dispersion effects [45]. The projector augmented wave (PAW) method [46, 47] was used to describe the interaction between the frozen core electrons (*i.e.* up to 3*p* for Ni, Cu, Zn and up to 1*s* for C, N, O) and its valence electrons and a kinetic energy cut-off of 400 eV was fixed for the plane-wave basis set expansion. A Γ-centred $k$-grid of 4×4×1 $k$-points was used, which leads to 6 irreducible reciprocal lattice points. During relaxation, the cell parameters are allowed to relax while keeping the cell volume constant, so that the vacuum gap is preserved. The forces on the atoms were minimised until they were less than 0.02 eV Å$^{-1}$.

All calculations were spin-polarised and we considered all possible spin states and relative orientations of the magnetic moments for the transition metal ions. For $Cu^{2+}$ ($d^9$), only one spin state is possible, with one unpaired electron. For $Ni^{2+}$ ($d^8$), two spin states are possible where there can be zero or two unpaired electrons, which correspond to low-spin (LS) or high-spin (HS), respectively. In the case of $Co^{2+}$ ($d^7$), two spin states are possible also where there can be one or three unpaired electrons (LS and HS, respectively). For the spin-polarised ions, both antiferromagnetic (AFM) and ferromagnetic (FM) configurations were considered.

In order to calculate more accurate electronic structures and band gap values of the materials, we performed single-point calculations on the GGA-optimised structures, using a screened hybrid functional (HSE06) [48, 49]. For these calculations, a reduced mesh of 2×2×1 $k$-point mesh was used. All the electron energies are reported with respect to the vacuum reference. As in other periodic DFT codes, the band energies in VASP are given with respect to an internal energy reference. Therefore, to obtain absolute energy levels it is necessary to evaluate the electrostatic potential in the pseudo-vacuum region represented by an empty space within the simulation cell, with zero potential gradient. This is chosen here as the planar average in middle of the vacuum gap between the nanosheets. The MacroDensity code was employed for this purpose [50].

Using the Gaussian16 code [51], time-dependent DFT calculations (TD-DFT) were also performed in order to examine the excited states in some selected cases. For these calculations, the MOF systems were represented by cluster models consisting of one porphyrin and one paddlewheel unit, where all the cleaved C bonds were saturated with H atoms (**figures**

**S1**, **S2** and **S3** in the SI). The geometries of the clusters were fixed to those obtained from the periodic calculations. For consistency, we used the same HSE06 functional as in the VASP calculations. Triplet states were used to describe the magnetic nature of the copper paddlewheel clusters. The calculations were all-electron (i.e. no pseudopotentials were employed) and a 6-311G(d,p) basis set was used to expand the wavefunctions.

## 3. Results and discussion

### 3.1. Framework geometry and magnetic ground states

We first discuss how the nature of the paddlewheel metal affects the geometric and electronic properties of the framework. **Table 1** summarises the relative stabilities and geometric parameters of the frameworks with different paddlewheel compositions (Co, Ni, Cu or Zn), spin state and magnetic ordering. The metal at the centre of the porphyrin was kept as Zn in all cases.

Except in the case of the non-magnetic $Zn^{2+}$ cations, we need to investigate different spin states and magnetic ordering for each paddlewheel composition. The Cu paddlewheel exhibits antiferromagnetic (AFM) coupling between the two neighbouring metal centres in the paddlewheel. This is consistent with both experimental measurements in Cu paddlewheel clusters [52-54] and with the theoretical study by Rodriguez-Fortea, *et al*. [55]. For Co, the preferred spin state of each metal cation is high-spin (HS), which involves a local magnetic moment $\mu = 3$ $\mu_B$ per Co(II) cation. This result agrees with the experimental determination by Pakula and Berry, who showed that Co(II) species are in high-spin state in Co paddlewheel units [56], although they found that the local magnetic moments were aligned antiferromagnetically within the paddlewheel, whereas in our calculation we found that the ferromagnetic alignment is more stable.

**Table 1. Relative stability and geometric parameters for different spin states and magnetic configurations of the 2D frameworks at each paddlewheel composition. μ is the local magnetic moment on each metal atom. *d*[*M-M*] is the distance between the metal atoms in the paddlewheel. Energies are obtained from HSE calculations at PBE+U geometries.**

| $M$ | Spin state | μ ($\mu_B$) | Magnetic configuration | Relative stability (eV) | $a$ (Å) | $d$ [$M$-$M$] (Å) |
|---|---|---|---|---|---|---|
| Co(II) | **HS** | **3** | AFM | + 0.25 | 16.75 | 2.45 |
|  |  |  | **FM** | **0** | **16.74** | **2.53** |
|  | LS | 1 | AFM | + 0.50 | 16.65 | 2.34 |
|  |  |  | FM | + 0.69 | 16.63 | 2.36 |
| Ni(II) | HS | 2 | AFM | + 1.23 | 16.68 | 2.44 |
|  |  |  | FM | +0.43 | 16.71 | 2.58 |
|  | **LS** | **0** | -- | **0** | **16.60** | **2.38** |
| Cu(II) | -- | 1 | **AFM** | **0** | **16.73** | **2.45** |
|  |  |  | FM | + 0.03 | 16.70 | 2.45 |
| Zn(II) | -- | 0 | -- | -- | 16.80 | 2.55 |

For the Ni case, our calculations give the low-spin state as the ground state, contrasting with the experimental measurements by Pang *et al.* [53] who found a high-spin antiferromagnetic ground state in Ni paddlewheel clusters. In general, the comparison between theoretically and experimentally determined electronic ground states is difficult for these systems, because our calculations refer to systems with coordinatively unsaturated metal centres, i.e. only the four equatorial carboxylate ligands are considered, whereas in most experimental situations the metal centres in the paddlewheel are saturated by additional linkers or solvent molecules, e.g. water [53] or ethanol [56]. If we perform our calculation for the Ni-paddlewheel system with water molecules added in axial position making each Ni centre penta-coordinated, then we find that the high-spin antiferromagnetic state is the most stable, followed by the high-spin ferromagnetic ground state, which in this case is only 0.06 eV higher in energy. In any case, we have found that the band alignment presented below does not change much with the nature of the magnetic ground state. The results below refer to the magnetic ground states found theoretically in this work for the paddlewheel units with tetra-coordinated metal centres.

The cell parameter of the Zn-paddlewheel framework (16.80 Å) can be compared with the experimental value obtained by X-ray diffraction for stacked 2D porphyrin-paddlewheel nanosheets with the same composition, which was 16.71 Å [57]. The discrepancy (0.5%) is small considering the approximations in the DFT simulation, both related to the exchange-correlation functional and to ignoring vibrational effects. Furthermore, the experimental value refers to the bulk material, whereas the simulation corresponds to a single-layer material. The variation of both the cell parameter and the *M-M* distance is consistent with the trend of ionic radii for $Co^{2+}$, $Ni^{2+}$, $Cu^{2+}$ and $Zn^{2+}$, which is not monotonous along the period but has a minimum value for $Ni^{2+}$ [58].

### 3.2. Effect of changing the paddlewheel metal on the band positioning

We now discuss the suitability of the band structure alignment for the photocatalysis of solar fuel synthesis from $H_2O$ or $CO_2$, as a function of the nature of the metal in the paddlewheel. For this analysis, we need to align the band edges with respect to the vacuum potential, to obtain their absolute positions, and compare with the redox potentials for the photocatalytic reaction. The valence and conduction bands must straddle the redox potentials for the given reaction. For example, for water-splitting, the valence band edge should be below the energy of the oxygen evolution reaction (OER):

$$H_2O \leftrightarrow 2H^+_{(aq)} + \frac{1}{2}O_{2(g)} + 2e^- \qquad (1)$$

while the conduction band edge should be above the energy of the hydrogen evolution reaction (HER):

$$2H^+_{(aq)} + 2e^- \leftrightarrow H_{2(g)} \qquad (2)$$

The difference between these redox potentials is 1.23 eV and therefore the band gap needs to be higher than this value. The optimal band gap for water-splitting is ~2 eV [59] to take into consideration loss mechanisms, *i.e.* via thermal energy. At pH = 0 in the vacuum scale, the potential value for OER is -5.67 eV and for HER is -4.44 eV. These potentials are shifted by $k_B T \times pH \times \ln 10$ (where $k_B$ is Boltzmann's constant) for systems at temperature $T$ and $pH > 0$. The redox potentials values for water splitting at a neutral pH and room temperature are -5.26 eV and -4.03 eV for OER ($O_2/H_2O$) and HER ($H^+/H_2$), respectively. For $CO_2$ reduction the band edges must straddle a larger redox potential difference with carbon dioxide reduction to methane ($CO_2/CH_4$) at -3.79 eV and carbon dioxide reduction to methanol ($CO_2/CH_3OH$) at -3.65 eV. All these potentials and the positions of the band edges and band gaps are shown in **Figure 2**.

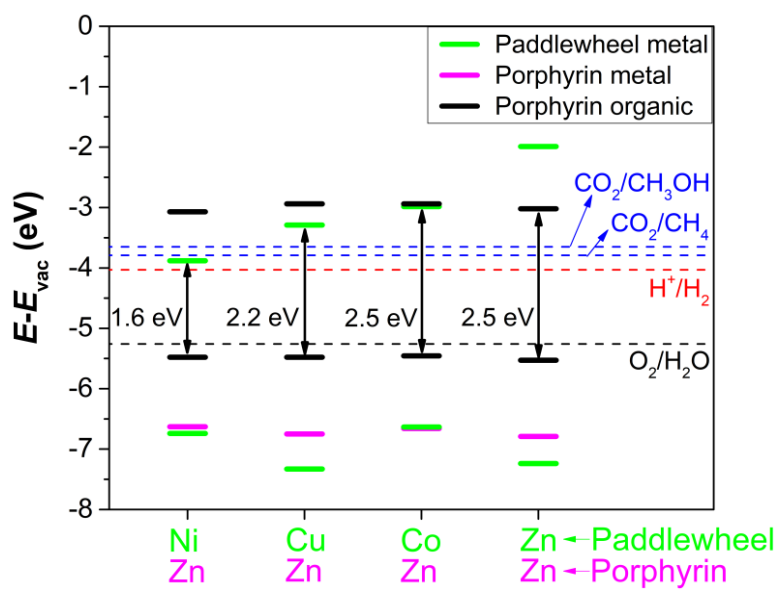

Figure 2. Band alignment of *M*-Zn-TCPP systems (*M* = $Co^{2+}$, $Ni^{2+}$, $Cu^{2+}$ and $Zn^{2+}$ is the metal in the paddlewheel). Levels from the paddlewheel metal, the porphyrin metal centre, and the organic part (C, N and H atoms) of porphyrins are shown in green, magenta and black, respectively. The energy levels of relevant half-reactions involved in water splitting and $CO_2$ reduction to $CH_4$ and $CH_3OH$ are also shown.

It is shown that the valence band, which is mainly contributed by the porphyrin unit, is approximately at -5.5 eV, which is below the OER as required. The nature of the paddlewheel metal mainly affects the position of the conduction band edge. Whereas the lowest unoccupied molecular orbital (LUMO) of the porphyrin is always at the same energy (~-3 eV), the empty 3*d* levels of the paddlewheel transition metal centres can go below that value, lowering the band gap. For the Ni-Zn system, a lower-lying, empty Ni 3*d* level narrows the band gap to 1.6 eV, where in the Co-Zn system, the lowest empty Co $3d$ level is at the same energy as the porphyrin's LUMO, so the band gap is not narrowed. For water-splitting photocatalysis, the best metal in the paddlewheel is Cu, whose empty $3d$ levels bring the band gap to 2.2 eV. The Co-Zn system also has suitable band positions, albeit with a larger gap, which might be useful for photocatalytic $CO_2$ reduction reactions.

Taking the Cu-Zn system as reference, we can form a picture of how the photocatalytic water-splitting reaction could occur in a system like this. A water molecule would interact with the porphyrin unit (we have calculated an adsorption energy of -0.26 eV for water at the Zn-porphyrin in this MOF), where a photogenerated hole would drive the water oxidation reaction, evolving oxygen gas. A ligand-to-metal charge transfer (LMCT) would then have to take place, moving the excited electron from the excited levels of the porphyrin to the metal paddlewheel, where the reduction of protons would occur, driven by the high energy of the excited electron in the Cu $3d$ state.

TD-DFT calculations in a cluster model of the Cu-Zn-TCPP system confirm that the lowest-energy electronic excitation ($T_1$) involves charge transfer from the porphyrin to the paddlewheel (Supplementary Information: see **Table S1** for the list of excited states and **Table S2** for the relative charges of the porphyrin and paddlewheel units). However, this first excitation has zero oscillator strength, i.e. the charge transfer cannot be achieved via direct excitation. The lowest bright excitation ($T_{44}$) is a transition localized within the porphyrin unit, and corresponds to the so-called Soret band (or B band) of the porphyrin, which typically appears in the far visible or ultraviolet (UV) region of the spectrum [21, 60-62]. These calculations suggest two limitations of these porphyrin-based MOFs in photocatalytic applications. First, most of the adsorption happens at energies in the far visible or UV range of the spectrum, so it would not be possible to take advantage of most of the energy from solar radiation, which lies in the visible region. Second, since the oscillator strength of the charge transfer state is very low, we need to engineer the structure to make charge transfer more feasible. Therefore, in the next sections we will consider possible modifications to these MOFs, which could enhance their photocatalytic properties.

**3.3. Effect of changing the metal from Zn to Co at the porphyrin centre**

We now briefly consider the substitution of Zn by Co at the porphyrin centre, for which we have investigated the band positions in a system with Co at the centre of the porphyrin and Cu in the paddlewheel. The conduction band for this Cu-Co system is 0.3 eV below the LUMO of the porphyrin, which is similar to what is observed for the Cu-Zn system (**Figure 3**).

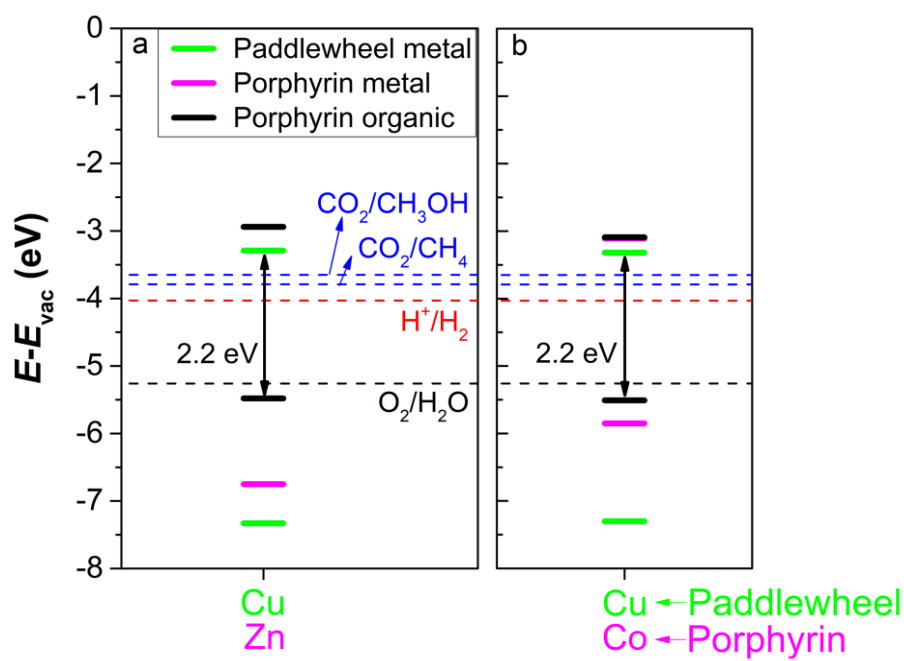

**Figure 3. Comparison between band alignments of a) Cu-Zn-TCPP and b) Cu-Co-TCPP systems. The energy levels of the metal at paddlewheel, the metal at the porphyrin centre and organic part (C, N and H atoms) of porphyrins are shown in green, magenta and black, respectively. The energy levels of relevant half-reactions involved in water splitting and $CO_2$ reduction to $CH_4$ and $CH_3OH$ are also shown.**

However, the highest filled 3$d$ levels of the Co centres are significantly higher in energy than the highest filled 3$d$ levels of Zn. The proximity between the high-lying filled Co 3$d$ levels and the HOMO of the porphyrin will help stabilize a photogenerated hole, since the Co(II) centre can be readily oxidised to Co(III) (in contrast with the case of Zn, whose low-lying filled 3*d* levels prevent the oxidation). The presence of Co as a redox centre in porphyrins has been widely used in the experimental design of porphyrin-based photocatalysts [63, 64].

Although Co-porphyrins seem more interesting for photocatalysis than Zn- porphyrins, in what follows we will consider other modifications of the porphyrin-based MOFs, while keeping the Zn at the porphyrin centre, for ease of calculations (the Co centres introduce additional magnetic degrees of freedom). The effect of other metal centres at the porphyrin on the electronic structure of porphyrin-based MOFs has been investigated in more detail in Ref. [29].

## 3.4. Effect of partially reducing the porphyrin unit to chlorin

Until this point, we have explored substitutions of the metals at the paddlewheel and porphyrin units. Another route to modify the electronic properties and optical behaviour of these MOFs is to partially reduce the porphyrin unit to form chlorin. Chlorins fall in the generic class of hydroporphyrins, that is, porphyrin derivatives in which one or more bonds are saturated by addition of hydrogen, resembling a photosynthetic chromophore found in nature [65]. It is known that chlorin exhibits stronger light absorption at lower energies compared to porphyrin [66]. Lu *et al.* have successfully synthesised chlorin-based MOFs that improve the photophysical properties of porphyrin-based MOFs for photodynamic therapy of colon cancers [67]. The motivation here is to study the increase of the light absorption of the MOF in the visible range to improve its photocatalytic efficiency.

The Cu-Zn-TCPP system was used as starting point, where one of the pyrrole rings was reduced by hydrogenation, as shown in **Figure 4**. After optimisation, this system exhibited negligible changes of cell parameters and geometry compared to the Cu-Zn-TCPP one. In this structure, because of the application of periodic boundary conditions, the position of the reduced pyrrole is ordered. However, using a 2×2×1 supercell, we have considered different relative positions of the reduced pyrrole in neighbouring chlorin units, and found that all configurations have similar energies (within 0.1 meV) and band gaps (within 0.05 eV). Therefore, our analysis below refers to the ordered configuration represented by a single unit cell.

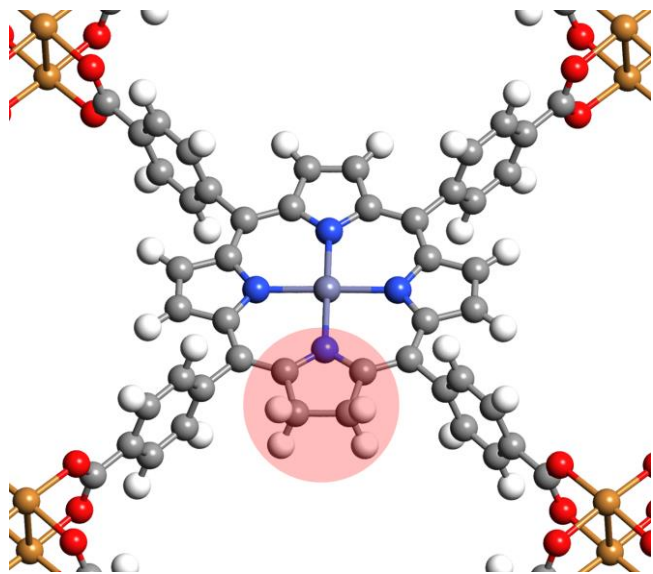

**Figure 4. The Cu-Zn-chlorin system: the $C_\beta$ atoms of the pink-shaded pyrrole are reduced. Colour code: grey = carbon, red = oxygen, white = hydrogen, blue = nitrogen, orange = copper.**

The partial reduction leads to a narrower band gap (1.9 eV), but otherwise the electronic structure is similar to that of the unreduced Cu-Zn-TCPP system (**Figure 5**). Although the valence band is now slightly above the OER level, it is possible in practical applications to realign such small differences using a bias voltage [29, 30].

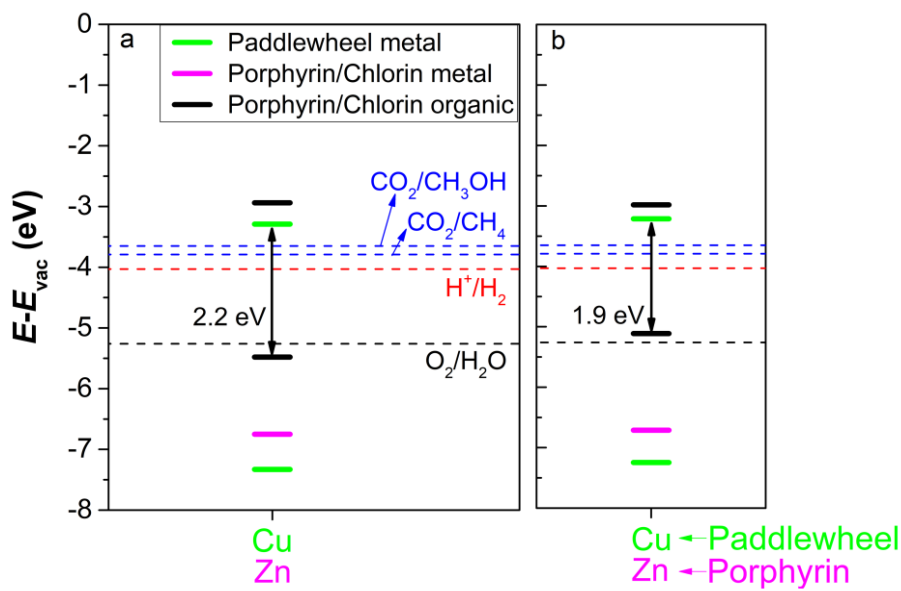

**Figure 5.** Comparison between band alignments of a) Cu-Zn-TCPP and b) Cu-Zn-chlorin systems. The bands of metal at paddlewheel, metal at the centre and organic part (C, N and H atoms of porphyrins) are shown in green, magenta and black, respectively. The energy levels of relevant half-reactions involved in water splitting and $CO_2$ reduction to $CH_4$ and $CH_3OH$ are also shown.

Excited states calculations were performed for these Cu-Zn-chlorin systems (energies of all calculated states and Bader charges of selected states are listed in **Tables S3** and **S4** of the SI). The first excited state ($T_1$), as in the unmodified porphyrin system, is a charge transfer state, but with zero oscillator strength. The Soret band is also found at roughly the same energy in the porphyrin- and chlorin-based MOFs. However, the chlorin-based MOF has a relatively bright state at lower energies ($T_{16}$ at 2.3 eV) which is not present in the porphyrin-based system. This is consistent with previous work showing that chlorin increases light absorption in the lower-energy Q bands, thus improving photophysical behaviour under visible light [66, 67]. This is clearly observed in the absorption spectra of the cluster model of the Cu-Zn-chlorin system (**Figure 6**), which shows the presence of a peak in the visible range. This peak corresponds to the Q bands, which in the case of the unreduced Cu-Zn-TCPP system is still appreciable but it has a very small intensity.

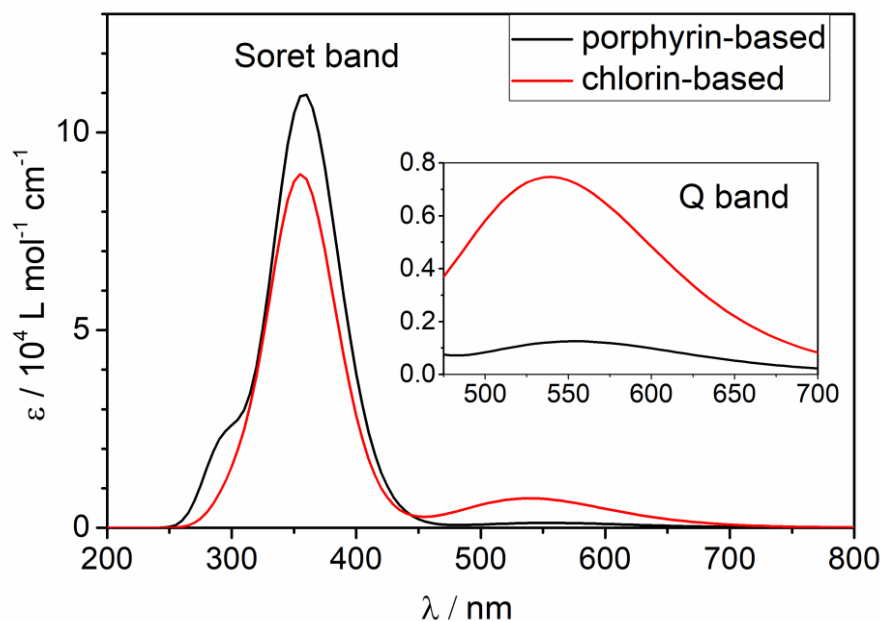

**Figure 6.** Light absorption spectra calculated using TD-DFT for the cluster models of the porphyrin-based (Cu-Zn-TCPP) and chlorin-based (Cu-Zn-chlorin) systems. The inset expands the spectrum in the region of the Q band.

All these results suggest that the partial reduction of porphyrin to chlorin units in these and other MOFs could enhance the photocatalytic performance under visible light, but to the best of our knowledge this avenue has not been experimentally explored.

**3.5. Changing the bridge between the porphyrin and paddlewheel**

Finally, we consider the effect of modifying the bridging species between the porphyrin and the paddlewheel units. We investigate the substitution of the benzene rings by ethyne (C2) or butadiyne (C4) bridges (**Figure 7**). We have shown in previous work that it is possible to modify the properties of porphyrin-based structures by varying the nature of the bridging species linking the porphyrins, and in this way tune the band gap values [68].

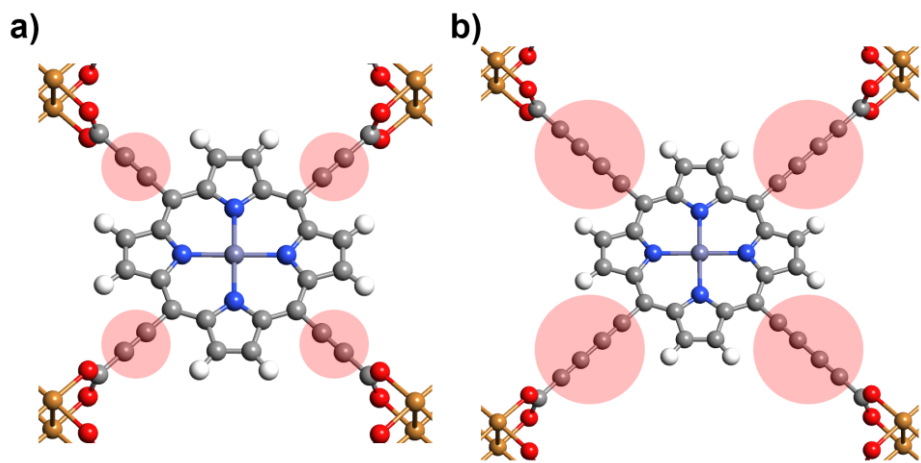

**Figure 7.** The two bridge-substituted systems: a) Cu-Zn-C2 and b) Cu-Zn-C4. The benzene rings have been substituted by the ethyne and butadiyne bridges, respectively, highlighted by pink-shaded circles. Colour code: grey = carbon, red = oxygen, white = hydrogen, blue = nitrogen, orange = copper.

The substitution of the benzene rings by C2 or C4 bridges induce significant narrowing of the band gaps (with values of 1.8 eV and 1.7 eV for C2 and C4, respectively) compared with the 2.2 eV value for the Cu-Zn-TCPP system (**Figure 8**). The gap narrowing is achieved mainly as consequence of the lowering of the conduction bands, since the valence bands are almost unaltered, going only slightly down in energy.

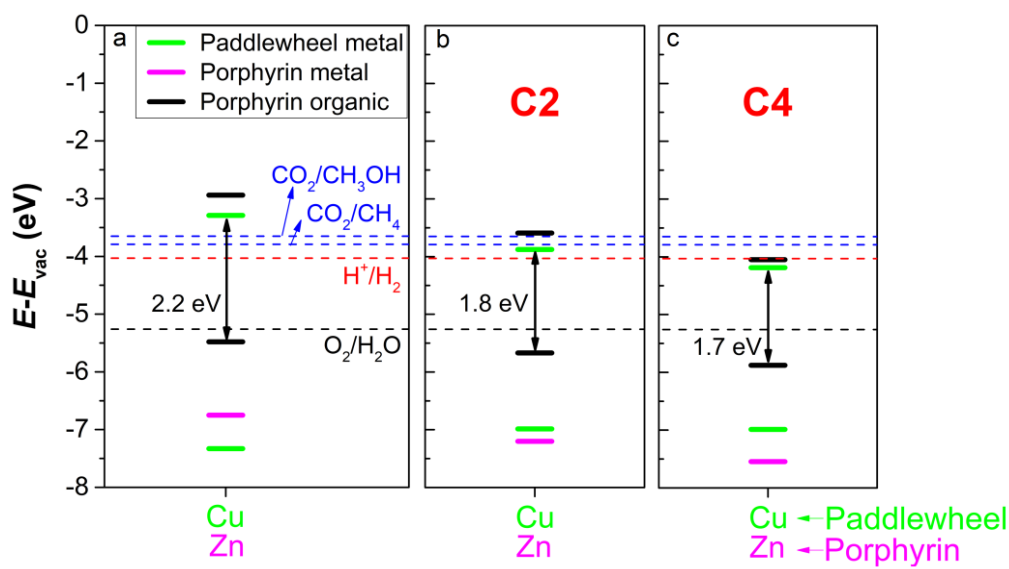

**Figure 8.** Comparison between band alignments of Cu-Zn-TCPP (left), Cu-Zn-C2 (middle) and -C4 (right) systems. The bands of metal at paddlewheel, metal at the centre and organic atoms of porphyrins (C, N and H) are shown in green, magenta and black, respectively. The energy levels of relevant half-reactions involved in water splitting and $CO_2$ reduction to $CH_4$ and $CH_3OH$ are also shown.

It is interesting to discuss the effect of the different bridges on the charge transfer. The excited states for the system with the C2 bridge show that there are now several relatively bright states involving charge transfer from the porphyrin to the paddlewheel (see Table S5 and Table S6 in the SI). Bright states $T_{30}$, $T_{34}$ and $T_{36}$, which are at energies in the region of the Soret band, now involve significant charge transfer from the porphyrin to the paddlewheel. This effect would lead to direct charge separation via light absorption and could potentially be beneficial for the photocatalytic behaviour of the system. Modifying the nature and length of the bridging units between the porphyrin and the metal cluster is thus the key to tune the charge transfer behaviour.

The length of the bridging unit length can also be expected to modify LMCT behaviour from excited states localized in the porphyrin. To illustrate this, we present here a simple model based on Marcus theory [69, 70], where we consider the porphyrin and the paddlewheel as two separate fragments. When light is absorbed at the bright states of the porphyrin, before any charge transfer takes place, the excitation will decay to the first excited state of the porphyrin (Kasha's rule). We make use Marcus theory to describe the electron transfer between the excited porphyrin fragment (donor) and the paddlewheel fragment (acceptor). To consider the electrostatic interaction between the charged fragments (porphyrin$^+$ and paddlewheel$^-$), we have simply added a classical electrostatic term to the energy of the charge transfer state, considering the distance between these fragments (and assuming a full electron transfer).

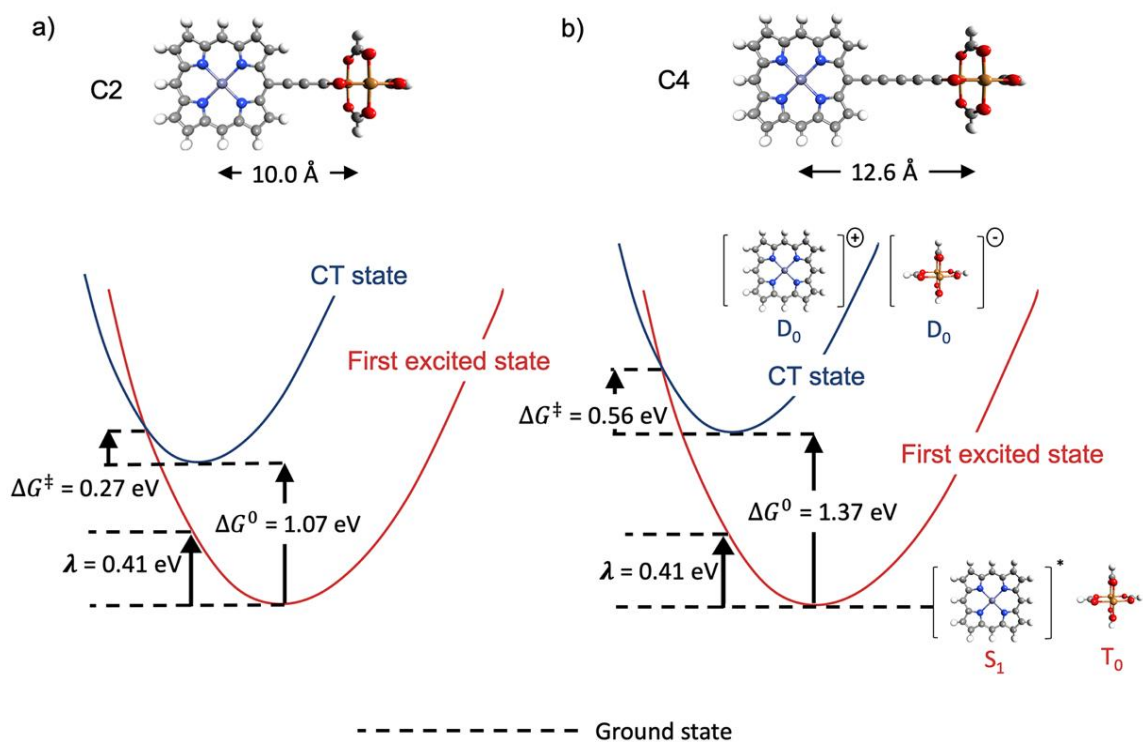

**Figure 9. Marcus parabolas and energy levels showing the transfer process in a) the C2 linker and b) the C4 linker. The excited state is shown in red and the charge transfer state in blue.**

In our model calculations, we find that the charge transfer state is always above the first excited state ($\Delta G^0 > 0$). In this case, the barrier (with respect to the charge transfer state) is given by $\Delta G^\ddagger = (\Delta G^0 - \lambda)^2/4\lambda$, where $\lambda$ is the reorganization energy (the change in energy of the first excited state when moved to the geometry of the charge transfer state, which in this simple model is independent of the length of the bridging unit). The energy of the charge-transfer state, and then the value of $\Delta G^0$, is 0.3 eV lower for the C2 linker than that for the C4 linker, due to the stronger electrostatic attraction for the shorter bridge. This means that in the C2 system the kinetic barrier to go from state to the other will also be lower, as shown in **Figure 9**. Our analysis suggests that the introduction of shorter bridging units between the porphyrin and the metal clusters should lower the kinetic barriers for the LMCT process. This is consistent with the experimental work from Lan *et al.* [26], who concluded that electron transfer from the photoexcited porphyrins to Ru clusters was facilitated by the proximity of the Ru cluster to the porphyrin in their MOF.

## 4. Conclusions

We have presented a computer simulation study of two-dimensional, porphyrin(chlorin)-based MOFs, consisting of metalloporphyrin units (with $Zn^{2+}$ or $Co^{2+}$ at the centre) and metal paddlewheel clusters (made $Co^{2+}$, $Ni^{2+}$, $Cu^{2+}$ or $Zn^{2+}$), bridged by different organic linear chains. Our results illustrate several ways in which the electronic and optical properties of these MOFs can be modified by the different composition or structural degrees of freedom. Changing the metal in the paddlewheel cluster from Zn to Co, Cu, or Ni, lowers the position of the empty 3$d$ levels in that order. For Cu and Ni the empty 3d levels are slightly below the LUMO of the porphyrin, which is convenient to achieve charge separation via linker to metal charge transfer (LCMT). The copper paddlewheel seems ideal for water splitting photocatalysis in terms of the resulting band gap and relative position of empty levels. Changing the metal at the centre of the porphyrin from Zn to Co does not change of the band edges or the bandgap, but creates an easy redox centre capable of stabilising the photogenerated hole at the porphyrin.

Perhaps even more important than engineering the band alignment and gap, is to enhance the visible light absorption and to facilitate charge separation in the framework. To address the first issue, we consider the partial reduction of porphyrin to chlorin in the MOF, which we demonstrate leads to stronger absorption in the visible range of the spectrum. Charge separation, on the other hand, is found to be sensitive to the nature and length of the bridging units between the porphyrin and the paddlewheel. For example, using an ethyne bridge introduces charge-transfer bright states in the energy region of the Soret band. Finally, we used a simple model based on Marcus theory to illustrate that the energy barriers to the LMCT process can be expected to decrease with the length of the bridge. Overall, our study suggests different avenues to modify porphyrin-based MOFs to improve their photocatalytic performance in applications such as water splitting and $CO_2$ reduction.

**Acknowledgements**

V.P. acknowledges a PhD studentship from the SENESCYT agency in Ecuador. This work made use of ARCHER, the UK's national high-performance computing service, via the UK's HPC Materials Chemistry Consortium, which is funded by EPSRC (EP/R029431), and of the Young supercomputer, via UK Materials and Molecular Modelling Hub, which is partially funded by EPSRD (EP/T022213/1).

# Supplementary Information

# Engineering the electronic and optical properties of 2D porphyrin-paddlewheel metal-organic frameworks

Victor Posligua,[1] Dimpy Pandya,[1] Alex Aziz,[2] Miguel Rivera,[2] Rachel Crespo-Otero,[2] Said Hamad,[3] Ricardo Grau-Crespo[1*]

[1]*Department of Chemistry, University of Reading, Whiteknights, Reading RG6 6AD, United Kingdom.*

*Email: r.grau-crespo@reading.ac.uk*

[2]*School of Biological and Chemical Sciences, Queen Mary University of London, Mile End Road, London E1 4NS, United Kingdom.*

[3]*Department of Physical, Chemical and Natural Systems, Universidad Pablo de Olavide, Ctra.de Utrera km.1, 41013 Seville, Spain.*

**List of content**

# 1. Excited states of Cu-paddlewheel / Zn-porphyrin system

The model we have employed for the TD-DFT calculation of excited states in the Cu-Zn-TCPP system is shown in **Figure S1**.

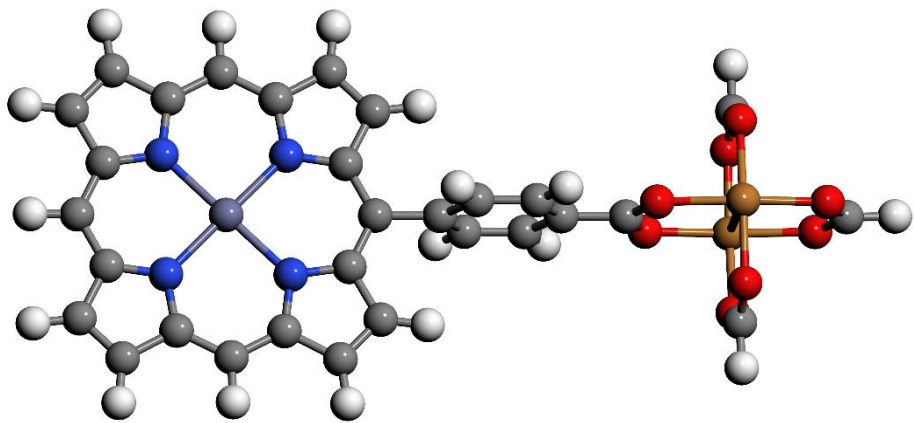

**Figure S1. Molecular model of Cu-Zn-porphyrin system. Colour code: grey = carbon, red = oxygen, white = hydrogen, blue = nitrogen, orange = copper.**

The excited states energies and oscillator strengths obtained for this model are presented in **Table S2**, and the atomic (Bader) charges in **Table S3**.

**Table S2. Excited states and oscillator strengths calculated for the Cu-Zn-TCPP system. Bright states are highlighted in orange.**

| States | Energy (eV) | Oscillator strength |
|---|---|---|
| $T_1$ | 1.43 | 0.00 |
| $T_2$ | 1.50 | 0.00 |
| $T_3$ | 1.61 | 0.00 |
| $T_4$ | 1.62 | 0.00 |
| $T_5$ | 1.86 | 0.00 |
| $T_6$ | 1.91 | 0.00 |
| $T_7$ | 1.93 | 0.00 |
| $T_8$ | 2.01 | 0.00 |
| $T_9$ | 2.01 | 0.00 |
| $T_{10}$ | 2.01 | 0.00 |
| $T_{11}$ | 2.02 | 0.00 |
| $T_{12}$ | 2.24 | 0.02 |
| $T_{13}$ | 2.24 | 0.01 |
| $T_{14}$ | 2.28 | 0.00 |
| $T_{15}$ | 2.28 | 0.00 |
| $T_{16}$ | 2.29 | 0.00 |
| $T_{17}$ | 2.38 | 0.00 |
| $T_{18}$ | 2.38 | 0.00 |
| $T_{19}$ | 2.86 | 0.00 |
| $T_{20}$ | 2.86 | 0.00 |
| $T_{21}$ | 2.89 | 0.00 |
| $T_{22}$ | 2.90 | 0.00 |
| $T_{23}$ | 3.08 | 0.00 |
| $T_{24}$ | 3.14 | 0.00 |
| $T_{25}$ | 3.15 | 0.00 |
| $T_{26}$ | 3.16 | 0.00 |
| $T_{27}$ | 3.17 | 0.00 |
| $T_{28}$ | 3.22 | 0.00 |
| $T_{29}$ | 3.24 | 0.00 |
| $T_{30}$ | 3.24 | 0.00 |
| $T_{31}$ | 3.26 | 0.00 |
| $T_{32}$ | 3.27 | 0.00 |
| $T_{33}$ | 3.28 | 0.00 |
| $T_{34}$ | 3.28 | 0.00 |
| $T_{35}$ | 3.29 | 0.00 |
| $T_{36}$ | 3.29 | 0.00 |
| $T_{37}$ | 3.31 | 0.00 |
| $T_{38}$ | 3.33 | 0.00 |
| $T_{39}$ | 3.33 | 0.00 |
| $T_{40}$ | 3.36 | 0.00 |
| $T_{41}$ | 3.41 | 0.00 |
| $T_{42}$ | 3.42 | 0.00 |
| $T_{43}$ | 3.44 | 0.00 |
| $T_{44}$ | 3.44 | 1.47 |
| $T_{45}$ | 3.45 | 0.00 |
| $T_{46}$ | 3.46 | 0.83 |
| $T_{47}$ | 3.56 | 0.00 |
| $T_{48}$ | 3.57 | 0.00 |
| $T_{49}$ | 3.57 | 0.00 |
| $T_{50}$ | 3.61 | 0.00 |
| $T_{51}$ | 3.64 | 0.00 |
| $T_{52}$ | 3.66 | 0.00 |
| $T_{53}$ | 3.70 | 0.00 |
| $T_{54}$ | 3.70 | 0.00 |
| $T_{55}$ | 3.73 | 0.01 |
| $T_{56}$ | 3.74 | 0.00 |
| $T_{57}$ | 3.76 | 0.00 |
| $T_{58}$ | 3.80 | 0.07 |
| $T_{59}$ | 3.81 | 0.03 |
| $T_{60}$ | 3.81 | 0.03 |
| $T_{61}$ | 3.84 | 0.02 |
| $T_{62}$ | 3.86 | 0.00 |
| $T_{63}$ | 3.87 | 0.02 |
| $T_{64}$ | 3.89 | 0.00 |
| $T_{65}$ | 3.98 | 0.00 |
| $T_{66}$ | 4.00 | 0.00 |
| T$_{67}$ | 4.08 | 0.00 |
| T$_{68}$ | 4.09 | 0.00 |
| T$_{69}$ | 4.11 | 0.00 |
| T$_{70}$ | 4.14 | 0.00 |
| T$_{71}$ | 4.18 | 0.00 |
| T$_{72}$ | 4.19 | 0.00 |
| T$_{73}$ | 4.20 | 0.00 |
| T$_{74}$ | 4.21 | 0.16 |
| T$_{75}$ | 4.22 | 0.30 |

**Table S3.** Bader charges of the Cu-Zn-TCPP system, calculated for the paddlewheel, bridge and porphyrin units ($\Delta Q_{PW}$, $\Delta Q_{bridge}$ and $\Delta Q_{porph}$, respectively), for the ground state, first excited state, and the two brightest states (highlighted in orange). The oscillator strengths are also shown.

| States | $\Delta Q_{PW}$ | $\Delta Q_{bridge}$ | $\Delta Q_{porph}$ | Oscillator strength |
|---|---|---|---|---|
| T$_0$ | 0 | 0 | 0 | - |
| T$_1$ | -0.72 | -0.04 | 0.76 | 0.00 |
| T$_{44}$ | 0 | 0 | 0 | 1.47 |
| T$_{46}$ | 0 | 0 | 0 | 0.83 |

## 2. Excited states after partial reduction of porphyrin to chlorin

The structural model used is shown in **Figure S2**. The excited states energies and oscillator strengths obtained for this model are presented in **Table S3** and the Bader charges are in **Table S4**.

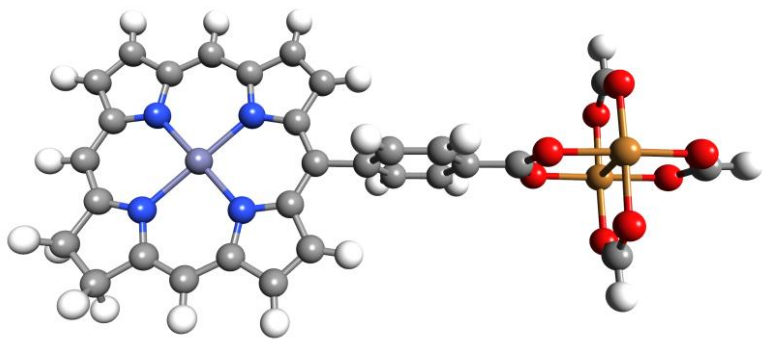

**Figure S2. Molecular model of Cu-Zn-Chlorin system. Colour code: grey = carbon, red = oxygen, white = hydrogen, blue = nitrogen, orange = copper.**

**Table S4. Excited states and oscillator strengths calculated for the Cu-Zn-Chlorin system. Bright states are highlighted in orange.**

| States | Energy (eV) | Oscillator strength |
|---|---|---|
| $T_1$ | 1.10 | 0.00 |
| $T_2$ | 1.43 | 0.00 |
| $T_3$ | 1.44 | 0.00 |
| $T_4$ | 1.52 | 0.00 |
| $T_5$ | 1.61 | 0.00 |
| $T_6$ | 1.86 | 0.00 |
| $T_7$ | 1.91 | 0.00 |
| $T_8$ | 2.01 | 0.00 |
| $T_9$ | 2.01 | 0.00 |
| $T_{10}$ | 2.03 | 0.00 |
| $T_{11}$ | 2.24 | 0.02 |
| $T_{12}$ | 2.24 | 0.01 |
| $T_{13}$ | 2.28 | 0.00 |
| $T_{14}$ | 2.28 | 0.00 |
| $T_{15}$ | 2.29 | 0.00 |
| $T_{16}$ | 2.31 | 0.13 |
| $T_{17}$ | 2.36 | 0.00 |
| $T_{18}$ | 2.46 | 0.01 |
| $T_{19}$ | 2.71 | 0.00 |
| $T_{20}$ | 2.77 | 0.00 |
| $T_{21}$ | 2.93 | 0.00 |
| $T_{22}$ | 2.96 | 0.00 |
| $T_{23}$ | 2.98 | 0.00 |
| $T_{24}$ | 3.02 | 0.00 |
| $T_{25}$ | 3.02 | 0.00 |
| $T_{26}$ | 3.06 | 0.00 |
| $T_{27}$ | 3.10 | 0.00 |
| $T_{28}$ | 3.13 | 0.00 |
| $T_{29}$ | 3.18 | 0.00 |
| $T_{30}$ | 3.19 | 0.00 |
| $T_{31}$ | 3.28 | 0.00 |
| $T_{32}$ | 3.29 | 0.00 |
| $T_{33}$ | 3.29 | 0.00 |
| $T_{34}$ | 3.31 | 0.00 |
| $T_{35}$ | 3.35 | 0.00 |
| $T_{36}$ | 3.39 | 0.00 |
| $T_{37}$ | 3.39 | 0.00 |
| $T_{38}$ | 3.43 | 0.01 |
| $T_{39}$ | 3.43 | 0.00 |
| $T_{40}$ | 3.44 | 0.00 |
| $T_{41}$ | 3.45 | 0.86 |
| $T_{42}$ | 3.48 | 0.88 |
| $T_{43}$ | 3.53 | 0.00 |
| $T_{44}$ | 3.57 | 0.00 |
| $T_{45}$ | 3.60 | 0.00 |
| $T_{46}$ | 3.60 | 0.11 |
| $T_{47}$ | 3.61 | 0.00 |
| $T_{48}$ | 3.62 | 0.00 |
| $T_{49}$ | 3.62 | 0.00 |
| $T_{50}$ | 3.64 | 0.04 |
| $T_{51}$ | 3.78 | 0.00 |
| $T_{52}$ | 3.82 | 0.00 |
| $T_{53}$ | 3.84 | 0.04 |
| $T_{54}$ | 3.86 | 0.00 |
| $T_{55}$ | 3.87 | 0.02 |
| $T_{56}$ | 3.94 | 0.22 |
| $T_{57}$ | 4.02 | 0.00 |
| $T_{58}$ | 4.03 | 0.00 |
| $T_{59}$ | 4.04 | 0.00 |
| $T_{60}$ | 4.05 | 0.00 |
| $T_{61}$ | 4.07 | 0.00 |
| $T_{62}$ | 4.08 | 0.00 |
| $T_{63}$ | 4.09 | 0.00 |
| $T_{64}$ | 4.10 | 0.00 |
| $T_{65}$ | 4.10 | 0.01 |
| $T_{66}$ | 4.11 | 0.00 |
| $T_{67}$ | 4.14 | 0.00 |
| $T_{68}$ | 4.14 | 0.00 |
| T$_{69}$ | 4.14 | 0.05 |
| T$_{70}$ | 4.15 | 0.00 |
| T$_{71}$ | 4.17 | 0.00 |
| T$_{72}$ | 4.17 | 0.00 |
| T$_{73}$ | 4.23 | 0.03 |
| T$_{74}$ | 4.25 | 0.00 |
| T$_{75}$ | 4.27 | 0.00 |

**Table S5. Bader charges of the Cu-Zn-Chlorin system, calculated for the paddlewheel, bridge and porphyrin units (ΔQ$_{PW}$, ΔQ$_{bridge}$ and ΔQ$_{porph}$, respectively), for the ground state, first excited state, and the two brightest states (highlighted in orange). The oscillator strengths are also shown.**

| States | ΔQ$_{PW}$ | ΔQ$_{bridge}$ | ΔQ$_{porph}$ | Oscillator strength |
|---|---|---|---|---|
| T$_0$ | 0 | 0 | 0 | |
| T$_1$ | -0.83 | 0.05 | 0.78 | 0.00 |
| T$_{16}$ | 0 | -0.01 | 0.01 | 0.13 |
| T$_{41}$ | -0.01 | 0.11 | -0.10 | 0.86 |

## 3. Excited states of system with C2 bridge

The model for the system with a C2 bridge between the porphyrin and the paddlewheel is shown in **Figure S3**. The excited states energies and oscillator strengths are presented in **Table S6**, and the Bader charges in **Table S7**.

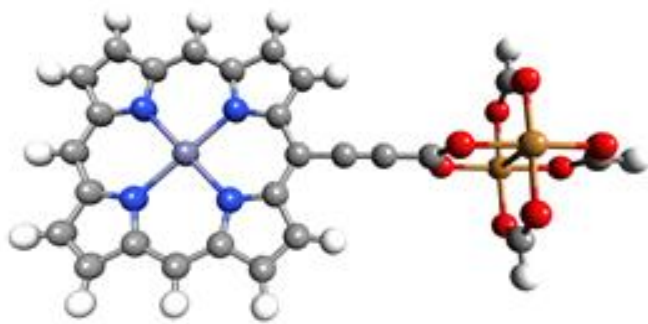

**Figure S3. Molecular model for the system where the benzene bridge is substituted by an ethyne (C2) bridge. Colour code: grey = carbon, red = oxygen, white = hydrogen, blue = nitrogen, orange = copper.**

**Table S6. Excited states and oscillator strengths calculated for the Cu-Zn-C2 system. Bright states are highlighted in orange.**

| States | Energy (eV) | Oscillator strength |
|---|---|---|
| $T_1$ | 1.40 | 0.04 |
| $T_2$ | 1.54 | 0.00 |
| $T_3$ | 1.57 | 0.06 |
| $T_4$ | 1.65 | 0.00 |
| $T_5$ | 1.89 | 0.00 |
| $T_6$ | 1.93 | 0.00 |
| $T_7$ | 1.95 | 0.00 |
| $T_8$ | 2.00 | 0.00 |
| $T_9$ | 2.03 | 0.00 |
| $T_{10}$ | 2.04 | 0.00 |
| $T_{11}$ | 2.08 | 0.00 |
| $T_{12}$ | 2.24 | 0.02 |
| $T_{13}$ | 2.24 | 0.01 |
| $T_{14}$ | 2.27 | 0.00 |
| $T_{15}$ | 2.28 | 0.00 |
| $T_{16}$ | 2.29 | 0.00 |
| $T_{17}$ | 2.31 | 0.00 |
| $T_{18}$ | 2.32 | 0.02 |
| $T_{19}$ | 2.86 | 0.03 |
| $T_{20}$ | 2.93 | 0.00 |
| $T_{21}$ | 2.99 | 0.00 |
| $T_{22}$ | 3.04 | 0.00 |
| $T_{23}$ | 3.04 | 0.00 |
| $T_{24}$ | 3.05 | 0.05 |
| $T_{25}$ | 3.12 | 0.00 |
| $T_{26}$ | 3.15 | 0.00 |
| $T_{27}$ | 3.17 | 0.00 |
| $T_{28}$ | 3.18 | 0.00 |
| $T_{29}$ | 3.22 | 0.03 |
| $T_{30}$ | 3.27 | 0.66 |
| $T_{31}$ | 3.30 | 0.00 |
| $T_{32}$ | 3.32 | 0.02 |
| $T_{33}$ | 3.34 | 0.73 |
| $T_{34}$ | 3.34 | 0.10 |
| $T_{35}$ | 3.37 | 0.00 |
| $T_{36}$ | 3.41 | 0.56 |
| $T_{37}$ | 3.44 | 0.00 |
| $T_{38}$ | 3.46 | 0.00 |
| $T_{39}$ | 3.46 | 0.00 |
| $T_{40}$ | 3.49 | 0.00 |
| $T_{41}$ | 3.53 | 0.00 |
| $T_{42}$ | 3.54 | 0.00 |
| $T_{43}$ | 3.55 | 0.00 |
| $T_{44}$ | 3.57 | 0.00 |
| $T_{45}$ | 3.64 | 0.00 |
| $T_{46}$ | 3.66 | 0.05 |
| $T_{47}$ | 3.66 | 0.00 |
| $T_{48}$ | 3.67 | 0.00 |
| $T_{49}$ | 3.67 | 0.00 |
| $T_{50}$ | 3.68 | 0.05 |
| $T_{51}$ | 3.69 | 0.00 |
| $T_{52}$ | 3.70 | 0.00 |
| $T_{53}$ | 3.74 | 0.00 |
| $T_{54}$ | 3.75 | 0.00 |
| $T_{55}$ | 3.79 | 0.00 |
| $T_{56}$ | 3.80 | 0.00 |
| $T_{57}$ | 3.82 | 0.04 |
| $T_{58}$ | 3.83 | 0.01 |
| $T_{59}$ | 3.86 | 0.02 |
| $T_{60}$ | 3.90 | 0.02 |
| $T_{61}$ | 3.90 | 0.00 |
| $T_{62}$ | 3.93 | 0.00 |
| $T_{63}$ | 3.93 | 0.00 |
| $T_{64}$ | 3.94 | 0.01 |
| $T_{65}$ | 3.96 | 0.01 |
| T$_{66}$ | 4.09 | 0.00 |
| T$_{67}$ | 4.09 | 0.22 |
| T$_{68}$ | 4.10 | 0.00 |
| T$_{69}$ | 4.16 | 0.00 |
| T$_{70}$ | 4.16 | 0.00 |
| T$_{71}$ | 4.21 | 0.00 |
| T$_{72}$ | 4.22 | 0.11 |
| T$_{73}$ | 4.26 | 0.00 |
| T$_{74}$ | 4.26 | 0.01 |
| T$_{75}$ | 4.27 | 0.00 |

**Table S7. Bader charges of the Cu-Zn-C2 system, calculated for the paddlewheel, bridge and porphyrin units ($\Delta Q_{PW}$, $\Delta Q_{bridge}$ and $\Delta Q_{porph}$, respectively), for the ground state and some relevant excited states. The oscillator strengths are also shown.**

| States | $\Delta Q_{PW}$ | $\Delta Q_{bridge}$ | $\Delta Q_{porph}$ | Oscillator strength |
|---|---|---|---|---|
| T$_0$ | 0 | 0 | 0 | - |
| T$_1$ | -0.31 | 0.03 | 0.28 | 0.04 |
| T$_3$ | -0.33 | 0.03 | 0.30 | 0.06 |
| T$_{30}$ | -0.34 | 0.05 | 0.29 | 0.66 |
| T$_{33}$ | 0 | 0.03 | -0.03 | 0.73 |
| T$_{34}$ | -0.48 | -0.11 | 0.59 | 0.10 |
| T$_{36}$ | -0.23 | 0.06 | 0.17 | 0.56 |